\documentclass[sigconf, screen]{acmart}

\usepackage[utf8]{inputenc}
\usepackage[T1]{fontenc} 

\usepackage{natbib}

\usepackage[framemethod=tikz]{mdframed}
\usetikzlibrary{shadows}
\newmdenv[tikzsetting={draw=gray,fill=white},
          roundcorner=5pt,shadow=true]{mdboxshad}
\mdfsetup{%
middlelinewidth=1.5pt
}

\usepackage{algorithmic}
\usepackage{algorithm}
\usepackage{array}
\newcolumntype{H}[1]{>{\raggedright\arraybackslash\hyphenpenalty=10000}p{#1}}
\usepackage[caption=false,font=normalsize,labelfont=sf,textfont=sf]{subfig}
\usepackage{textcomp}
\usepackage{stfloats}
\usepackage{amsmath}
\usepackage{url}
\usepackage{verbatim}
\usepackage{graphicx}

\makeatletter
\def\zifour@spc{\hyphenchar\font=`\-\relax}
\makeatother
\usepackage{hyperref}
\usepackage{csquotes}
\usepackage{enumitem}
\setlist[itemize,enumerate]{leftmargin=7mm}

\usepackage{colortbl}
\usepackage{xcolor}
\usepackage{multirow}
\usepackage{tabularx}
\usepackage{booktabs}
\usepackage{comment}
\usepackage[super]{nth}
\usepackage{hyperref}

\definecolor{background-gray}{gray}{0.96}
\usepackage[framemethod=tikz]{mdframed}
\newmdenv[
  topline=false,
  bottomline=false,
  rightline=false,
  nobreak=true,
  skipabove=\baselineskip,
  skipbelow=\baselineskip,
  innertopmargin=6pt,
  innerbottommargin=6pt,
  innerleftmargin=12pt,
  innerrightmargin=12pt,
  tikzsetting={draw=black, line width=4pt},
  linecolor=black,
  backgroundcolor=background-gray
]{results}

\newcommand{\boxedtext}[1]{\fbox{\scriptsize\bfseries\textsf{#1}}}
\newcommand{\nota}[2]{
	\boxedtext{#1}{\small$\blacktriangleright$\emph{\textsl{#2}}$\blacktriangleleft$}
}

\newcommand{\todo}[1]{{\color{red}\nota{TODO}{#1}}} 
\renewcommand\todo[1]{}

\newcommand*{\enq}[1]{\enquote{{\itshape#1}}}

\setcopyright{cc}
\setcctype{by}
\acmDOI{10.1145/3805760.3814887}
\acmYear{2026}
\copyrightyear{2026}
\acmISBN{979-8-4007-2601-9/2026/07}
\acmConference[AIware '26]{Proceedings of the 3rd ACM International Conference on AI-Powered Software}{July 6--7, 2026}{Montreal, QC, Canada}
\acmBooktitle{Proceedings of the 3rd ACM International Conference on AI-Powered Software (AIware '26), July 6--7, 2026, Montreal, QC, Canada}
\acmSubmissionID{fseaiware26main-pp062-p}
\received{2026-02-15}
\received[accepted]{2026-03-28}

\begin{document}

\title[Harness Engineering for Agentic AI Coding Tools]{Harness Engineering for Agentic AI Coding Tools: An~Exploratory~Study\titlenote{This article is an extended version of \emph{Configuring Agentic AI Coding Tools: An Exploratory Study}, published at AIware~2026~\cite{galster2026configuring}.}}

\author{Matthias Galster}
\affiliation{%
 \institution{University of Bamberg}
 \city{Bamberg}
 \country{Germany}
}
\email{mgalster@ieee.org}
\orcid{0000-0003-3491-1833}

\author{Seyedmoein Mohsenimofidi}
\affiliation{%
  \institution{Heidelberg University}
  \city{Heidelberg}
  \country{Germany}
}
\email{s.mohsenimofidi@uni-heidelberg.de}
\orcid{0009-0009-1620-2735}

\author{Jai Lal Lulla}
\affiliation{%
  \institution{Singapore Management University}
  \city{Singapore}
  \country{Singapore}}
\email{jailal.l.2025@phdcs.smu.edu.sg}
\orcid{0009-0005-0024-8238}

\author{Muhammad Auwal Abubakar}
\affiliation{%
  \institution{University of Bamberg}
  \city{Bamberg}
  \country{Germany}
}
\email{muhammad.abubakar@uni-bamberg.de}
\orcid{0009-0006-1028-0650}

\author{Christoph Treude}
\affiliation{%
  \institution{Singapore Management University}
  \city{Singapore}
  \country{Singapore}
}
\email{ctreude@smu.edu.sg}
\orcid{0000-0002-6919-2149}

\author{Sebastian Baltes}
\affiliation{%
  \institution{Heidelberg University}
  \city{Heidelberg}
  \country{Germany}
}
\email{sebastian.baltes@uni-heidelberg.de}
\orcid{0000-0002-2442-7522}

\renewcommand{\shortauthors}{Galster et al.}

\begin{abstract}
Agentic AI coding tools increasingly automate software development tasks.
Developers can configure these tools through versioned repository-level artifacts such as Markdown and JSON files.
The emerging term \emph{harness engineering} extends context engineering, broadening the focus from a model's context to the full set of mechanisms configured around it.
We present a systematic analysis of configuration mechanisms for agentic AI coding tools, covering Claude Code, GitHub Copilot, Cursor, Gemini, and Codex.
We identify eight configuration mechanisms spanning from static context to executable and external integrations and, in an empirical study of 2,853 GitHub repositories, examine whether and how they are adopted, with a detailed analysis of \textsc{Context Files}, \textsc{Skills}, and \textsc{Subagents}.
First, \textsc{Context Files} dominate the configuration landscape and are often the sole mechanism in a repository, with \texttt{AGENTS.\allowbreak{}md} emerging as an interoperable standard across tools.
Second, few repositories adopt advanced mechanisms such as \textsc{Skills} and \textsc{Subagents}. \textsc{Skills} predominantly rely on static instructions rather than executable scripts.
Third, distinct configuration practices are forming around different tools, with Claude Code users employing the broadest range of mechanisms.
These findings establish an empirical baseline for understanding harness engineering, suggest that \texttt{AGENTS.\allowbreak{}md} serves as a natural starting point, and motivate longitudinal and experimental research on how configuration strategies evolve and affect agent performance.
\end{abstract}



%




\maketitle

\setlength{\parfillskip}{0pt plus 0.75\columnwidth}

\section{Introduction}

Agentic AI coding tools 
based on large language models (LLMs)~\cite{SERGEYUK2025107610, 10.1145/3715003}, automate time-consuming and repetitive tasks, such as generating and editing code, tests, and documentation. 
Unlike purely reactive conversational assistants, these tools proactively accomplish defined objectives~\cite{anthropicClaudeCodeRelease} by autonomously interacting with development environments, project artifacts, external data, and command-line tools with reduced human interaction~\cite{yang2024swe}.
Certain functionality can be delegated to other \emph{tools} and \emph{agents}.
An \emph{agent} is a goal-directed component that interprets a user goal, decomposes it into substeps, selects and executes tools, and iteratively adjusts its plan in an \emph{agent loop}.
\emph{Tools} are deterministic capabilities with a specific, bounded function (e.g., a Bash tool to run shell commands\footnote{\url{https://code.claude.com/docs/en/tools-reference}}), invoked by the underlying model in an \emph{agent loop}.
We distinguish such fine-grained tools from broader \emph{agentic AI (coding) tools} such as Claude Code or OpenAI Codex.

Initially, these tools implemented one central agent loop steered by a foundation model.
Conversational tools such as GitHub Copilot and Cursor soon offered similar capabilities via an \emph{agent mode}.
More recently, tool vendors introduced extension and configuration mechanisms to customize tool behavior, one of which allows developers to define their own \emph{subagents} that operate in parallel to the central agent loop, in their own context.

We use \emph{context} to denote the complete input to a single model call.
The software layers around the model(s) that drive the agent loop (i.e., the agent \emph{harness}\footnote{\url{https://magazine.sebastianraschka.com/p/components-of-a-coding-agent}}) assemble this context for each call, expose tool schemas, and manage turn-by-turn state.
\emph{Context engineering}~\cite{DBLP:journals/corr/abs-2507-13334} is the practice of designing the context that the harness assembles at runtime.
Through various configuration mechanisms, developers customize the harness for a given project, and thereby the context the model sees on each call.
This customization of the harness, not only the context, is \emph{harness engineering}.\footnote{\url{https://martinfowler.com/articles/harness-engineering.html}}

We define a \emph{configuration mechanism} as a means for developers to tailor tool and agent behavior to a project or workflow (e.g., context files or dedicated subagent definitions). A \emph{configuration artifact} is a tangible instance of a mechanism: either a single configuration file (e.g., \texttt{CLAUDE.\allowbreak{}md}, or one subagent file such as \texttt{.\allowbreak{}claude/\allowbreak{}agents/\allowbreak{}reviewer.\allowbreak{}md}) or a directory bundling several configuration files that together define one artifact (e.g., a skill whose directory contains \texttt{SKILL.\allowbreak{}md} alongside scripts, references, and assets).
Context files such as \texttt{AGENTS.\allowbreak{}md} or \texttt{CLAUDE.\allowbreak{}md} act as ``READMEs for agents'' with context-specific information about build commands, coding conventions, and rules for CI/CD pipelines~\cite{Seyedmoein2026, openaiIntroducingCodex}. 
These configuration artifacts can be version-controlled, making them inspectable and collaboratively maintainable. 
With the increased availability and diversity of agentic tools, the number of available configuration mechanisms and related artifacts has increased.

To understand which configuration mechanisms agentic tools offer and how they are used in open-source software (OSS), we address the following \textbf{research questions:}

\begin{description}[style=multiline, labelindent=2mm, leftmargin=10mm, topsep=4pt]
\item[RQ1] \emph{What configuration mechanisms do agentic coding tools offer?}
\item[RQ2] \emph{Which mechanisms are adopted in OSS repositories?}
\item[RQ3] \emph{How are configuration mechanisms adopted?}
\end{description}

We restrict our scope to configuration mechanisms whose artifacts are consumed by agentic tools and are repository-versioned.

\looseness=-1
With this paper, we provide the following \textbf{contributions:}
(1) We systematically document eight configuration mechanisms: \textsc{Context Files}, \textsc{Skills}, \textsc{Subagents}, \textsc{Commands}, \textsc{Rules}, \textsc{Settings}, \textsc{Hooks}, and \textsc{MCP} servers. We identified these from the documentation of Claude Code, GitHub Copilot, Cursor CLI, Gemini CLI, and Codex CLI. Some mechanisms are available in all tools; others are specific to particular tools.
(2) We analyzed the adoption of these configuration mechanisms in 2,853 GitHub repositories. 
\textsc{Context Files} (Markdown files that provide contextual project information) dominated and are often the sole configuration mechanism. Claude Code users apply the broadest range of configuration mechanisms.
(3) We analyzed the adoption of \textsc{Context Files}, \textsc{Skills}, and \textsc{Subagents} in more detail.
Three \textsc{Context File} formats showed the most dynamic development, with \texttt{AGENTS.\allowbreak{}md} emerging as a standard. Although \textsc{Skills} and \textsc{Subagents} can be defined for a wide range of purposes, most repositories that adopt them define only one or two artifacts. Moreover, to extend agent behavior, \textsc{Skills} primarily rely on static resources rather than executable scripts.

\begin{figure*}[tb]
    \centering
    \includegraphics[width=1\linewidth]{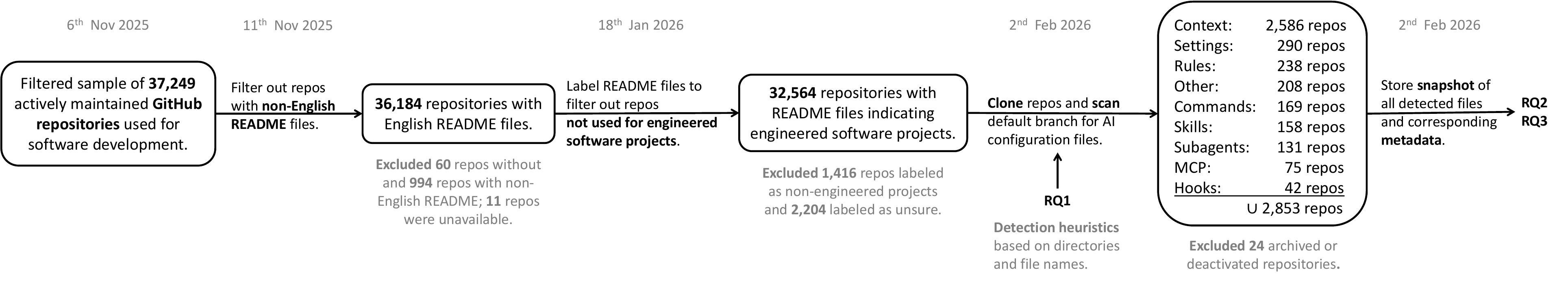}
    \caption{Data collection process.}
    \Description{Diagram showing the data collection process.}
    \label{fig:data-collection}
\end{figure*}

\section{Related Work}
\label{sec:related_work}

Prior work on context engineering has focused on single artifact types in isolation, without systematic examination of the broader configuration mechanisms.
Context engineering builds on earlier prompt engineering techniques~\cite{DBLP:journals/corr/abs-2402-07927, promptingguideElementsPrompt} but broadens the scope from individual prompts to the systematic design of all information provided to the model. \citeauthor{DBLP:journals/corr/abs-2507-13334}~\cite{DBLP:journals/corr/abs-2507-13334} surveyed context retrieval, processing, and management techniques for LLMs, while practitioner accounts demonstrate how structuring project-specific context improves agent effectiveness in complex codebases~\cite{philschmidSkillPrompting, dexterHorthyComplexCodebases}.
In a recent review of LLM agents, \citeauthor{DBLP:journals/corr/abs-2604-08224}~\cite{DBLP:journals/corr/abs-2604-08224} traced a progression from model weights to context to the harness, naming the last step \emph{harness engineering}.
Empirical work has mainly focused on repository-maintained context files, such as \texttt{AGENTS.\allowbreak{}md}. \citeauthor{Seyedmoein2026}~\cite{Seyedmoein2026} showed that repository-level context files describe project architecture, build commands, and contribution conventions; \citeauthor{DBLP:journals/corr/abs-2511-12884}~\cite{DBLP:journals/corr/abs-2511-12884} additionally found that they are actively maintained through small frequent changes. \citeauthor{santos2025decoding}~\cite{santos2025decoding} analyzed 328 \texttt{CLAUDE.\allowbreak{}md} files from Claude Code projects and found that architectural concerns dominate, followed by development guidelines, project overviews, and testing.

Beyond structural characterization, \citeauthor{lulla2026impact}~\cite{lulla2026impact} ran a controlled study with and without an \texttt{AGENTS.\allowbreak{}md} file, reporting lower runtime and token consumption at comparable task completion.
\citeauthor{villamizar2025prompts}~\cite{villamizar2025prompts} argued that prompts should be treated as software engineering artifacts and outlined a research agenda for their evolution, reuse, and governance. Q.~Zhang et al.~\cite{zhang2025agentic} proposed a framework treating contexts as evolving instructions, addressing brevity bias and context collapse through incremental updates.

Other work formalizes context as a programmable abstraction. Y.~Zhang et al.~\cite{zhang2025monadic} introduced Monadic Context Engineering, modeling context construction and transformation through monadic composition.
\citeauthor{mcmillan2026structured}~\cite{mcmillan2026structured} evaluated structured file-native context schemas and found that a multi-file organization and retrieval strategy affects agent accuracy more than the choice of serialization format. \citeauthor{ye2026meta}~\cite{ye2026meta} proposed Meta Context Engineering, a framework in which agentic skills and context artifacts co-evolve via evolutionary search. \citeauthor{Jiang2026}~\cite{Jiang2026} explored Cursor rules and developed a taxonomy of their content.

\citeauthor{DBLP:journals/corr/abs-2601-18341}~\cite{DBLP:journals/corr/abs-2601-18341} measured coding-agent adoption on GitHub through file-, commit-, and pull-request-level heuristics in 128,018 repositories. Their categorization distinguishes structured \emph{configuration files} (such as YAML) from natural-language \emph{rules and guidance files} (including \texttt{CLAUDE.\allowbreak{}md} and \texttt{AGENTS.\allowbreak{}md}).
This is a coarser split than our eight-mechanism taxonomy. Their sample applies activity and size thresholds without an engineered-project classification step, and their analysis centers on adoption rates and AI-assisted commit characteristics rather than on the internal structure and co-occurrence of configuration mechanisms.

However, no prior work maps configuration mechanisms across tools or studies their adoption patterns across repositories. We provide this cross-tool perspective.

\section{Data Collection and Analysis}
\label{sec:data_collection_analysis}

We collected OSS projects hosted on GitHub, selecting repositories belonging to ``engineered'' software projects~\cite{DBLP:journals/ese/MunaiahKCN17} using an updated version of a selection approach from prior work~\cite{Seyedmoein2026}.
The repository selection started with the SEART GitHub search tool~\cite{seartGitHubSearch, DBLP:conf/msr/DabicAB21}.
We applied several filters, each addressing a specific sampling risk~\cite{DBLP:conf/msr/KalliamvakouGBSGD14}.
We selected non-fork repositories with at least two contributors and a license, because forks duplicate codebases, the presence of a license signals the author's intent that the project be reused, and single-contributor projects often represent personal experiments rather than collaborative software.
The creation cutoff of {1}~January~2024 ensures that each repository predates the broad availability of agentic AI coding tools by more than a year, so detected configuration mechanisms represent additions to established projects rather than greenfield experiments built around the tools.
The activity cutoff of commits since {1}~June~2025 removes dormant projects and keeps repositories that were actively maintained during the period when agentic tools became widely available.
We excluded archived, disabled, or locked repositories for the same reason.
We applied a licensing filter that retained only OSI-approved software licenses and a popularity-based language filter, selecting the ten most common primary languages (Python, TypeScript, JavaScript, Go, Java, C++, Rust, PHP, C\#, and~C).
We further excluded repositories with fewer than 271 commits or fewer than 7 watchers, the median values of the sampling frame following \citeauthor{Seyedmoein2026}~\cite{Seyedmoein2026}, filtering out low-activity or poorly-followed repositories.
These filters resulted in a sample of 37,249 repositories.

Figure~\ref{fig:data-collection} outlines the data collection pipeline for this sample.
We cloned the repositories and searched their default branch for a \texttt{README} file, excluding 11 that became unavailable and 60 without such a file.
We then used the \texttt{lingua-language-detector} Python library to detect each file's language, excluding 994 repositories with non-English \texttt{README} files.

\begin{table*}
\centering
\footnotesize
\caption{Overview of repository-level configuration mechanisms across agentic AI coding tools. Each cell lists the repository file(s) or directory implementing that mechanism; ``--'' indicates that the mechanism was not available when we checked.}
\label{tab:config-options-overview}
\begin{tabular}{l>{\raggedright\arraybackslash}p{2.7cm}H{1.9cm}H{2.8cm}H{2.2cm}H{2.1cm}H{1.9cm}}
\toprule
\textbf{Mechanism} & \textbf{Description} & \textbf{Claude} & \textbf{Copilot} & \textbf{Codex} & \textbf{Cursor} & \textbf{Gemini} \\
\midrule
\textsc{Context Files} & Markdown file loaded into the context each session. & \texttt{CLAUDE.\allowbreak{}md} & \texttt{.github/~\{}\newline \texttt{copilot-instructions.md} | \texttt{instructions/*.md~\}}$^{a}$ & \texttt{AGENTS.\allowbreak{}md},\newline \texttt{AGENTS.\allowbreak{}override.\allowbreak{}md} & \texttt{AGENTS.\allowbreak{}md}, \texttt{.\allowbreak{}cursorrules}$^{c}$ & \texttt{GEMINI.\allowbreak{}md} \\
\addlinespace
\textsc{Settings} & JSON/TOML config for project-level tool behavior. & \texttt{.claude/ settings (local)?.json} & --$^{b}$ & \texttt{.\allowbreak{}codex/\allowbreak{} config.\allowbreak{}toml} & \texttt{.\allowbreak{}cursor/\allowbreak{} cli.\allowbreak{}json} & \texttt{.gemini/\{ settings.json} | \texttt{config.yaml \}} \\
\addlinespace
\textsc{Skills} & Reusable knowledge and invocable workflows. & \texttt{.\allowbreak{}claude/\allowbreak{}skills/\allowbreak{}} & \texttt{.\allowbreak{}github/\allowbreak{}skills/\allowbreak{}} & \texttt{.\allowbreak{}codex/\allowbreak{}skills/\allowbreak{}} & \texttt{.\allowbreak{}cursor/\allowbreak{}skills/\allowbreak{}} & \texttt{.\allowbreak{}gemini/\allowbreak{}skills/\allowbreak{}} \\
\addlinespace
\textsc{Subagents} & Specialized agents that operate in parallel to the central agent loop, in their own context. & \texttt{.\allowbreak{}claude/\allowbreak{}agents/\allowbreak{}} & \texttt{.\allowbreak{}github/\allowbreak{}agents/\allowbreak{}} & -- & \texttt{.\allowbreak{}cursor/\allowbreak{}agents/\allowbreak{}} & -- \\
\addlinespace
\textsc{Commands} & User-triggered shortcuts for predefined prompts. & \texttt{.\allowbreak{}claude/\allowbreak{} commands/\allowbreak{}} & -- & -- & \texttt{.\allowbreak{}cursor/\allowbreak{} commands/\allowbreak{}} & \texttt{.\allowbreak{}gemini/\allowbreak{} commands/\allowbreak{}} \\
\addlinespace
\textsc{Hooks} & Scripts executed at specific agent lifecycle points. & \texttt{.\allowbreak{}claude/\allowbreak{} settings.\allowbreak{}json} & \texttt{.\allowbreak{}github/\allowbreak{}hooks/\allowbreak{}*.\allowbreak{}json} & -- & \texttt{.\allowbreak{}cursor/\allowbreak{} hooks.\allowbreak{}json} & \texttt{.\allowbreak{}gemini/\allowbreak{} settings.\allowbreak{}json} \\
\addlinespace
\textsc{Rules} & System-level instructions to control agent behavior. & -- & -- & \texttt{.\allowbreak{}codex/\allowbreak{}rules/\allowbreak{}} & \texttt{.\allowbreak{}cursor/\allowbreak{}rules/\allowbreak{}} & -- \\
\addlinespace
\textsc{MCP} & External tool or data connections via the Model Context Protocol. & \texttt{.\allowbreak{}mcp.\allowbreak{}json} & --$^{b}$ & \texttt{.\allowbreak{}codex/\allowbreak{} config.\allowbreak{}toml} & \texttt{.\allowbreak{}cursor/\allowbreak{} mcp.\allowbreak{}json} & \texttt{.\allowbreak{}gemini/\allowbreak{} settings.\allowbreak{}json} \\
\bottomrule
\end{tabular}
\vskip 2pt
\begin{minipage}{\textwidth}
\footnotesize
\centering
$^{a}$ Copilot also supports \texttt{CLAUDE.\allowbreak{}md}, \texttt{AGENTS.\allowbreak{}md}, and \texttt{GEMINI.\allowbreak{}md}; $^{b}$ Configured via the GitHub web UI, not via files in the project repository;\newline $^{c}$ Cursor deprecated \texttt{.\allowbreak{}cursorrules} and now suggests using \texttt{AGENTS.\allowbreak{}md} instead.
\end{minipage}
\end{table*}

\begin{table}
\centering
\footnotesize
\caption{Release dates of agentic AI coding tools and repositories 
($n=2,853$; one repository can use multiple tools).}
\begin{tabular}{lp{3.7cm}r}
\toprule
\textbf{Agentic Tool}  & \textbf{Release (Month/Year)} & \textbf{\#Repositories} \\
\midrule
Claude Code &  02/2025 (CLI \& agents since release)  & 1,297\\
\midrule
\multirow{3}{*}{GitHub Copilot} & 10/2021 (Release)\newline 02/2025 (Copilot Agent Mode)\newline 09/2025 (Copilot CLI) & \multirow{3}{*}{957}\\
\midrule
\multirow{2}{*}{\texttt{AGENTS.\allowbreak{}md}$^{d}$} & 05/2025 (Adoption in Codex)\newline 08/2025 (Specification~\cite{agentsmd_spec}) & \multirow{2}{*}{493}\\
\midrule
\multirow{2}{*}{Cursor CLI} & 03/2023 (Release)\newline 06/2025 (Cursor Agents)\newline 08/2025 (Cursor CLI) & \multirow{2}{*}{327}\\
\midrule
Gemini CLI & 02/2024 (Release)\newline 05/2025 (Gemini Agent Mode)\newline 06/2025 (Gemini CLI) & 175\\
\midrule
Codex CLI & 04/2025 (CLI \& agents since release)  & 4\\
\bottomrule
\end{tabular}
\vskip 2pt
\begin{minipage}{\columnwidth}
\footnotesize
\centering
$^{d}$ Repositories using \texttt{AGENTS.\allowbreak{}md} without any tool-specific configuration artifact.
\end{minipage}
\label{tab:ai_tools}
\end{table}

\begin{figure}[t]
    \centering
    \includegraphics[width=\columnwidth]{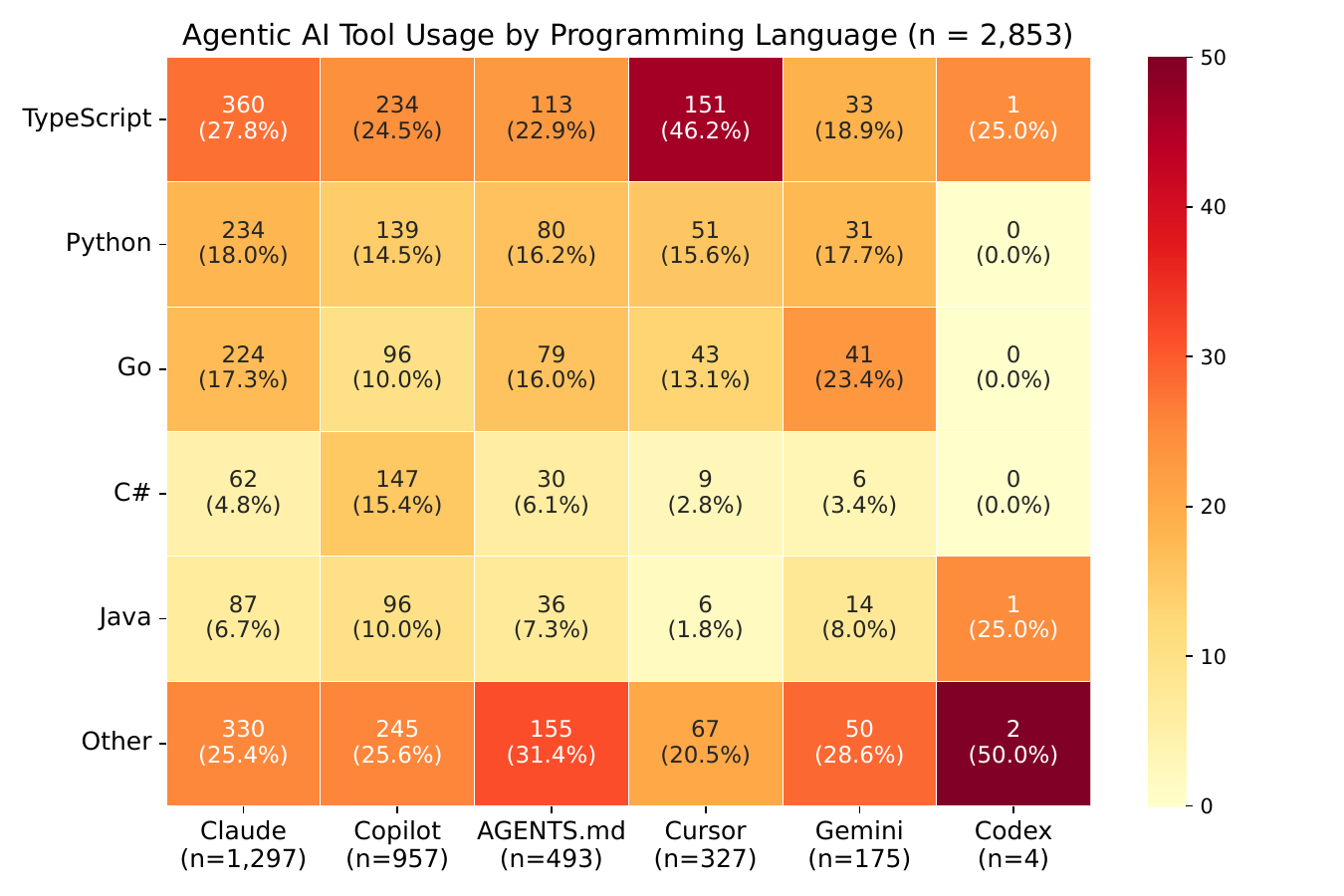}
    \caption{Adoption of agentic tools per programming language; \texttt{AGENTS.\allowbreak{}md} denotes repositories using only that file with no tool-specific configuration. Percentages relative to repository count per tool. Column totals can exceed the overall repository count (multiple tools per repository).}
    \Description{Diagram showing adoption of AI coding tools per programming language.}
    \label{fig:language-tool}
\end{figure}

\begin{figure}[t]
    \centering
    \includegraphics[width=\columnwidth]{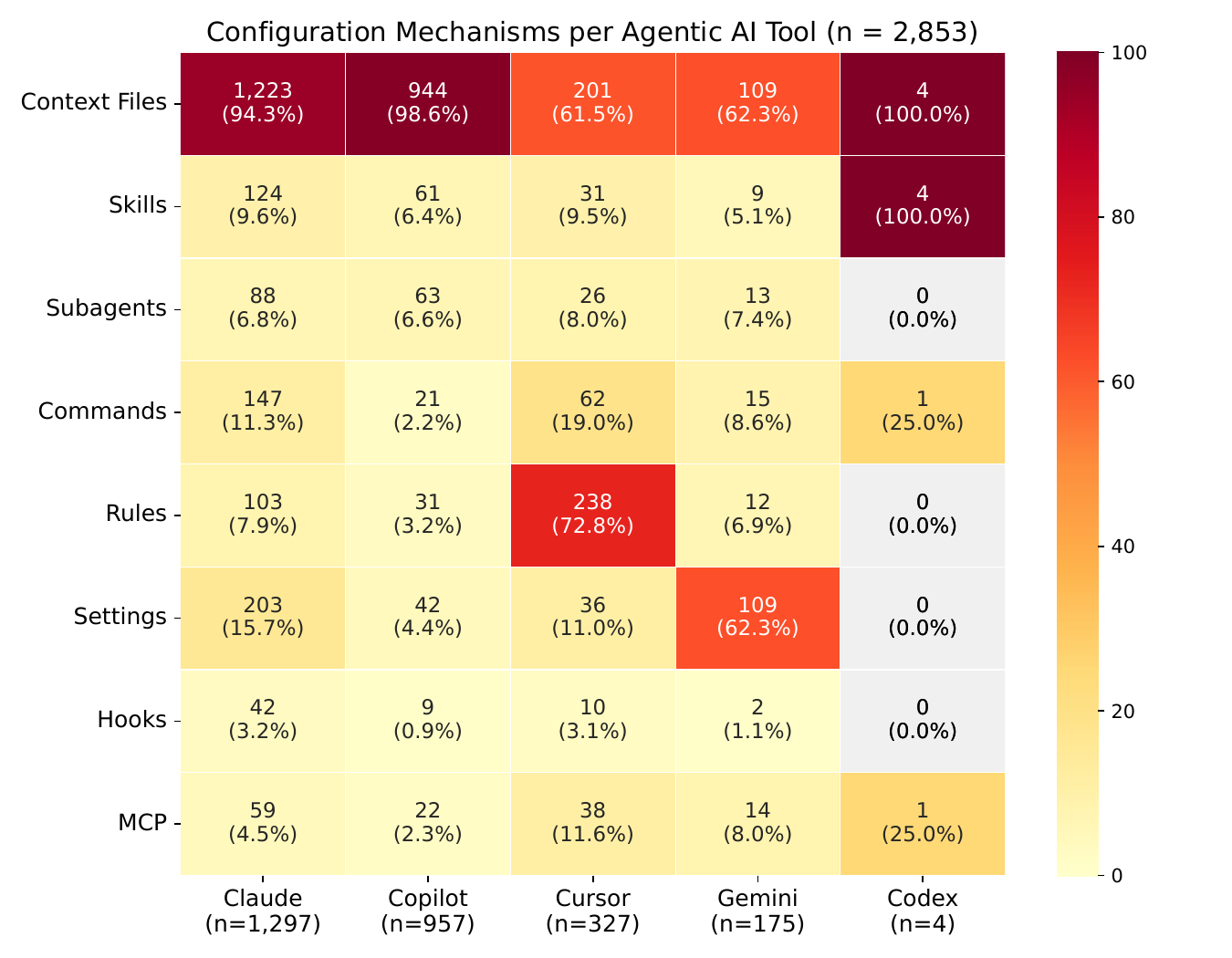}
    \caption{Usage of configuration mechanisms across agentic tools. Repositories using only \texttt{AGENTS.\allowbreak{}md} without tool-specific configuration are excluded. Percentages relative to repository count per tool. Column totals can exceed the overall repository count (multiple tools per repository).}
    \Description{Diagram showing usage of configuration mechanisms across agentic tools.}
    \label{fig:config-per-tool}
\end{figure}

For the remaining 36,184 repositories, we used a classification pipeline to determine, based on \texttt{README} content, whether each project qualifies as ``engineered'', that is, it shows clear evidence of software engineering practices~\cite{DBLP:journals/ese/MunaiahKCN17}. We operationalized this, among other criteria, as having a clear project purpose and documenting practices such as building, testing, and maintenance.
We used the OpenAI GPT-5.2 model for classification.
Our supplementary material includes the complete prompt, which we iteratively developed and tested on subsets of the sample, and the model configuration (we used the OpenAI defaults)~\cite{supplementary-material}.
The pipeline labeled 32,564 repositories as engineered and excluded the remaining 3,620 (1,416 classified as non-engineered, 2,204 `unsure').
We spot-checked randomly selected repositories from each category during prompt development and for the final sample.
We then cloned the remaining 32,564 repositories and applied heuristics based on file names and file paths to detect the usage of AI coding tools and configuration mechanisms (see Table~\ref{tab:config-options-overview}). These heuristics are based on the answers to \textbf{RQ1} and are briefly discussed in Section~\ref{sec:results_rq1}. This resulted in 2,853 repositories that use one or more AI coding tools; three repositories contained only empty configuration artifacts and were excluded.
Our data collection and analysis scripts and the analyzed data are available online~\cite{supplementary-material}.

\section{Configuration Mechanisms (RQ1)}
\label{sec:results_rq1}

We selected five agentic AI coding tools (Table~\ref{tab:ai_tools}) based on the \emph{2025 Stack Overflow Developer Survey}~\cite{stackoverflow-survey-2025}: the four most popular AI tools among the surveyed developers, with Codex substituting for ChatGPT (the agentic tool from the same vendor), plus Cursor, which has recently been studied in software engineering research~\cite{He2026, Jiang2026}.
Our study focuses on the CLI-based agentic interfaces that first appeared in 2025, although prior non-agentic versions exist for some tools.
One author systematically reviewed each tool's online documentation, documenting configuration mechanisms together with repository-level files and directories that indicate their usage.
Two other authors then cross-checked the extracted mechanisms and the heuristics developed to detect them, and all three authors discussed and reconciled the results.

Table~\ref{tab:config-options-overview} summarizes these heuristics. For example, a \texttt{.\allowbreak{}claude} directory or a \texttt{CLAUDE.\allowbreak{}md} file indicates Claude Code usage; a \texttt{.\allowbreak{}claude/\allowbreak{}agents/\allowbreak{}} directory with Markdown files indicates \textsc{Subagents}. \texttt{AGENTS.\allowbreak{}md} is a special case: as a tool-agnostic standard, it is supported by multiple tools, so repositories using only \mbox{\texttt{AGENTS.md}} without tool-specific artifacts are tracked separately (see Table~\ref{tab:ai_tools}, footnote~$^{d}$). Note that for Copilot, Cursor, and Gemini, certain detected files apply to both conversational and agentic workflows.
Our documentation of the mechanisms and detection heuristics (with links to the relevant tool documentation) and the Python scripts that implement our matching strategy are part of the supplementary material~\cite{supplementary-material}.

\begin{results}
\textbf{RQ1 (Summary):}
\begin{itemize}[leftmargin=*]
\item We identified eight configuration mechanisms spanning from static context (e.g., \textsc{Context Files}) to executable and external integrations (e.g., \textsc{Skills}, \textsc{MCP}). Two mechanisms (\textsc{Context Files} and \textsc{Skills}) are supported by all five tools.
\item Despite this convergence, no single tool implements all eight mechanisms.
\end{itemize}
\end{results}

\section{Adoption of Configuration Mechanisms (RQ2)}
\label{sec:results_rq2}

To understand how developers adopt configuration mechanisms in agentic AI coding tools, we analyzed the presence and co-occurrence of configuration artifacts across repositories.

\subsection{Characterization of Repositories in Dataset}
\label{subsec:characterization_of_repositories}

\looseness=-1
Before analyzing adoption, we characterize the 2,853 repositories in our dataset to contextualize the subsequent analysis.
Of the 2,853 repositories, 2,015 (70.6\%) adopted a single tool, while 295 (10.3\%) configured two tools, and 50 (1.8\%) configured three or more. An additional 493 repositories (17.3\%) used \texttt{AGENTS.\allowbreak{}md} as a tool-agnostic standard without any tool-specific configuration artifact; these are shown separately in the analysis (see Table~\ref{tab:ai_tools}, footnote~$^{d}$).
Among repositories with multiple tools, Claude appeared most frequently with others. The most common combination was Claude and Copilot ($n = 167$), followed by Claude and Cursor ($n = 144$), Claude and Gemini ($n = 52$), and Copilot and Cursor ($n = 49$). Notably, 44.0\% of Cursor repositories also configured Claude.
The dominance of single-tool repositories limits the potential for multi-tool confounding in our tool-specific analyses.
In our initial sample, the top five primary programming languages per repository ($n=36,184$) were Python (8,133; 22.5\%); TypeScript (4,999; 13.8\%); Java (3,813; 10.5\%); Go (3,786; 10.5\%); and JavaScript (3,709; 10.3\%). Interestingly, this order is slightly different for repositories that use agentic coding tools ($n=2,853$, see Figure~\ref{fig:language-tool}) where TypeScript and Go were more prominent and C\# replaces JavaScript in the top five: TypeScript (736, 25.8\%); Python (479, 16.8\%); Go (420, 14.7\%); C\# (228, 8.0\%); Java (223, 7.8\%).

\looseness=-1
TypeScript was the most common primary language for repositories using Claude, Copilot, and Cursor.
Usage in Java and C\# repositories was lower for all tools but Copilot.
As shown in Table~\ref{tab:repo-metadata-comparison}, which compares each tool's adopting repositories against the complement of non-adopting repositories, repositories using Cursor were younger than those using other tools. Excluding Codex ($n$=4), repositories associated with Gemini had larger contributor counts and commit volumes than those associated with other tools.
Cursor repositories were the largest by source code size (excluding Codex). \texttt{AGENTS.\allowbreak{}md} repositories tended to have fewer commits and smaller size than the overall sample, suggesting that these repositories adopted \texttt{AGENTS.\allowbreak{}md} as a lightweight, tool-agnostic starting point.

\subsection{Distribution of Configuration Mechanisms}

Figure~\ref{fig:config-per-tool} shows the distribution of configuration mechanisms per tool. \textsc{Context Files} (e.g., \texttt{CLAUDE.\allowbreak{}md}, \texttt{AGENTS.\allowbreak{}md}) were the most frequently adopted mechanism, used by 61.5 to 100\% of repositories across all tools.
Two tool-specific patterns stand out: 72.8\% of Cursor repositories adopt \textsc{Rules}, since Cursor was one of the first to introduce this mechanism~\cite{Jiang2026}, and 62.3\% of Gemini repositories use \textsc{Settings}. No other mechanism exceeds 20\% adoption for Claude, Copilot, Cursor, or Gemini; Codex ($n$=4) is too small to characterize reliably.

Most repositories included only a single \textsc{Context File} artifact, and the adoption of multiple non-context file mechanisms within the same repository remained rare. These results suggest that most repositories rely on \textsc{Context Files} as their baseline configuration.

\begin{figure*}[t]
    \centering
    \includegraphics[width=0.82\textwidth]{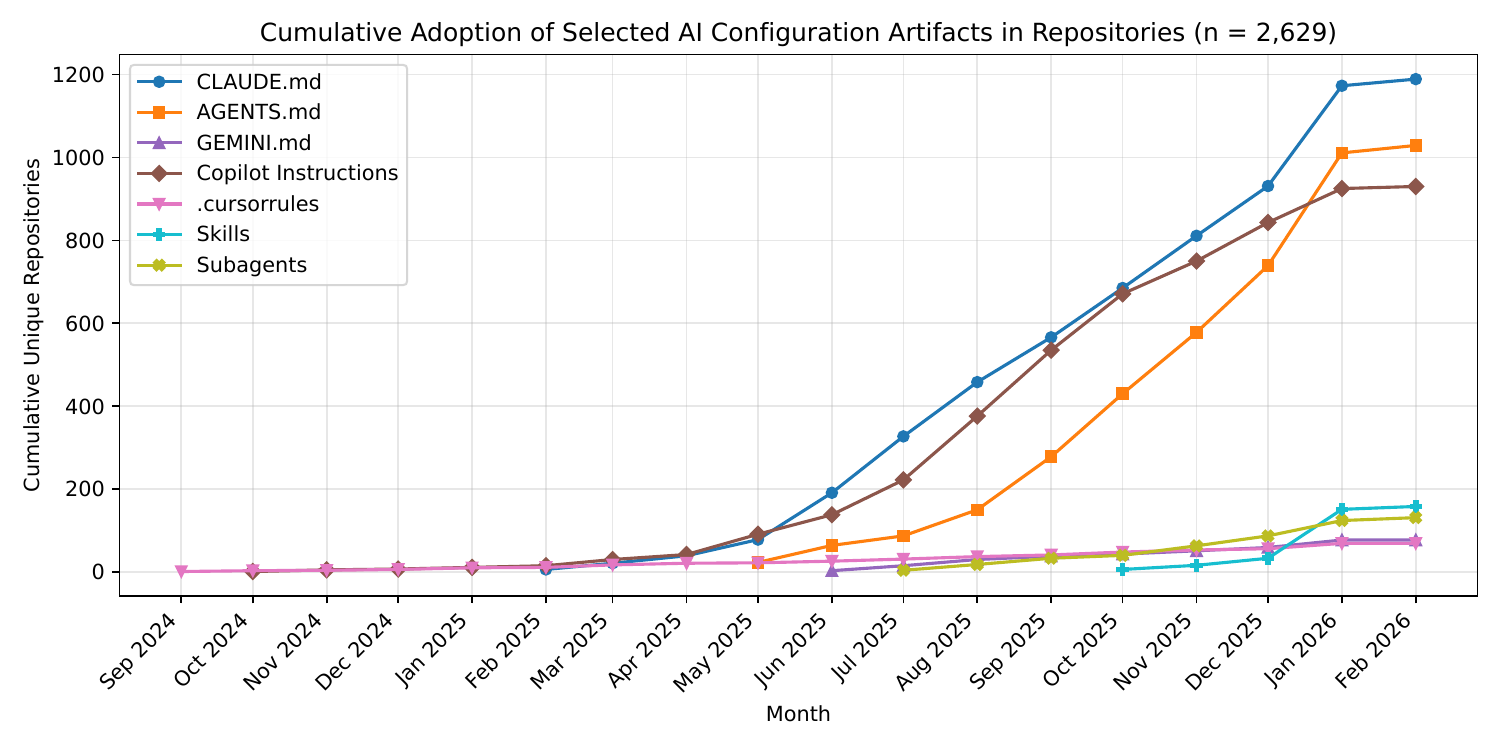}
    \caption{Cumulative adoption of selected configuration artifacts for agentic tools. Copilot Instructions comprise two \textsc{Context File} artifact types (file: \texttt{copilot-instructions.md}, directory: \texttt{instructions/*.md}, see Table~\ref{tab:config-options-overview}).} 
    \Description{Diagram showing the cumulative adoption of configuration artifacts for agentic tools.}
    \label{fig:adoption_over_time}
\end{figure*}

\begin{figure*}[t]
    \centering
    \includegraphics[width=0.68\textwidth]{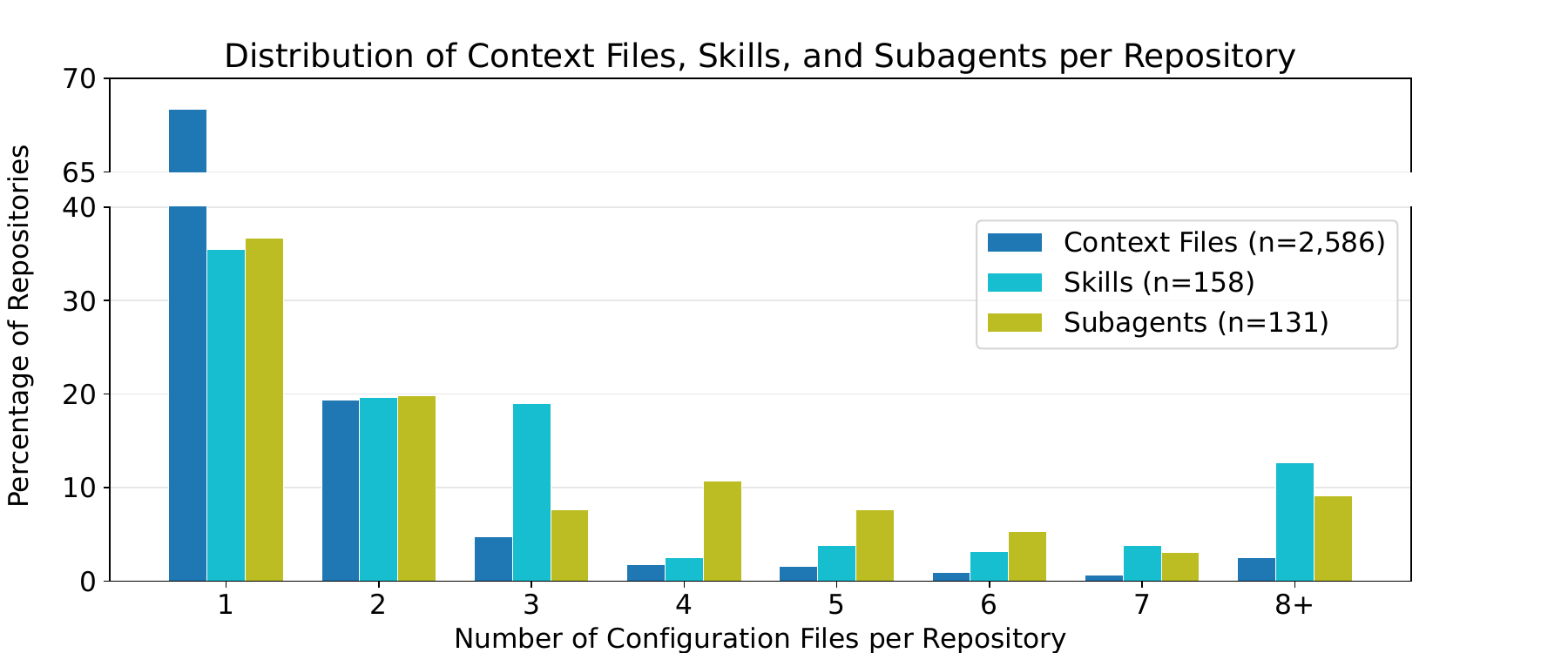}
    \caption{Configuration mechanism count per repository.}
    \Description{Diagram showing the configuration mechanism count per repository.}
    \label{fig:files-per-repo}
\end{figure*}

\subsubsection{Co-occurrence of Configuration Mechanisms}
We quantified pairwise associations between the adoption of configuration mechanisms using Cram\'{e}r's~$V$ from chi-squared tests (28~comparisons, Benjamini--Hochberg corrected).
Several configuration mechanisms are frequently adopted together, and some of these associations can be explained by shared configuration artifacts.
For example, \textsc{Settings} and \textsc{Hooks} show a positive association ($V = 0.36$), because hooks are usually defined in \textsc{Settings} files.
\textsc{Settings} and \textsc{Hooks} also both co-occur with \textsc{Skills} ($V = 0.15$ and $V = 0.24$, respectively), which are defined and configured differently.
\textsc{Subagents} also show positive associations with multiple mechanisms, most strongly with \textsc{Commands} ($V = 0.23$), \textsc{Skills} ($V = 0.15$), and \textsc{Hooks} ($V = 0.13$).
Other configuration mechanisms appear less frequently in combination. \textsc{Context Files} and \textsc{Rules} co-occur less often than independence would predict (Cram\'{e}r's $V = 0.41$; 122 observed vs.\ 216 expected). This under-representation can be attributed to how the mechanisms are defined and which tools first introduced them.
Overall, configuration mechanism adoption is characterized by recurring combinations rather than uniform usage, shaped by which mechanisms each tool supports (e.g., Claude, the dominating tool, does not support \textsc{Rules}). The full co-occurrence matrix is available in the supplementary material.

\subsubsection{Adoption of Configuration Mechanisms over Time}
Figure~\ref{fig:adoption_over_time} shows the adoption of individual \textsc{Context File} artifacts, \textsc{Skills}, and \textsc{Subagents} over time.
\textsc{Context Files} (e.g., \texttt{CLAUDE.\allowbreak{}md}, \texttt{AGENTS.\allowbreak{}md}) clearly dominate and have increased continuously, while \textsc{Skills} and \textsc{Subagents} experienced comparatively slow growth. \texttt{.\allowbreak{}cursorrules} and \texttt{copilot-instructions.\allowbreak{}md} started being introduced in 2024, although Cursor and Copilot were originally released in 2023 and 2021, respectively; their agentic capabilities were only introduced in 2025 (see Table~\ref{tab:ai_tools}). The adoption of Copilot instructions has increased since then.

\begin{results}
\textbf{RQ2 (Summary):}
\begin{itemize}[leftmargin=*]
\item \textsc{Context Files} are the dominant and often sole configuration mechanism; other mechanisms are adopted less frequently. Over time, this dominance has increased: \textsc{Skills} and \textsc{Subagents} have grown comparatively slowly.
\item Adoption varies: Claude Code repositories use the broadest range of mechanisms, while Cursor repositories emphasize \textsc{Rules}. Co-occurrence patterns are shaped by which mechanisms each tool supports.
\item Most repositories (70.6\%) adopt a single tool, and multi-tool overlap is limited. An additional 17.3\% use \texttt{AGENTS.\allowbreak{}md} without any tool-specific configuration.
\end{itemize}
\end{results}

\begin{table}
\centering
\footnotesize
\caption{Repository metadata by agentic tool. Cells show median, IQR, and Cliff's $\delta$ for tool-adopting repositories, each compared against the complement of non-adopting repositories. Significant differences (BH-adjusted) are in bold.}
\label{tab:repo-metadata-comparison}
\begin{tabular}{lrrrr}
\toprule
\textbf{Agentic tool} & \textbf{Age (years)} & \textbf{Contrib.} & \textbf{Commits} & \textbf{Size (KB)} \\
\midrule
All & 6.7 & 41 & 2,119 & 39k \\
($n$=2,853) & (4.3--9.4) & (19--104) & (965--5,073) & (9,280--132k) \\
\midrule
Claude & \textbf{6.2$^{***}$} & 40 & 2,216 & 41k \\
($n$=1,297) & \textbf{(4.1--9.2)} & (19--110) & (986--5,260) & (9,915--152k) \\
 & \textbf{$\delta$=-0.08} & $\delta$=0.01 & $\delta$=0.03 & $\delta$=0.04 \\
\addlinespace
Copilot & \textbf{7.1$^{***}$} & 42 & 2,231 & 45k \\
($n$=957) & \textbf{(5.0--9.7)} & (19--113) & (1,001--5,607) & (9,958--143k) \\
 & \textbf{$\delta$=0.11} & $\delta$=0.03 & $\delta$=0.04 & $\delta$=0.04 \\
\addlinespace
\texttt{AGENTS.\allowbreak{}md} & 6.9 & 42 & \textbf{1,841$^{**}$} & \textbf{24k$^{***}$} \\
($n$=493) & (4.2--9.5) & (18--88) & \textbf{(869--3,980)} & \textbf{(7,753--81k)} \\
 & $\delta$=0.03 & $\delta$=-0.04 & \textbf{$\delta$=-0.10} & \textbf{$\delta$=-0.13} \\
\addlinespace
Cursor & \textbf{5.5$^{***}$} & \textbf{52$^{*}$} & \textbf{2,780$^{***}$} & \textbf{75k$^{***}$} \\
($n$=327) & \textbf{(3.4--7.9)} & \textbf{(23--118)} & \textbf{(1,264--6,029)} & \textbf{(21k--198k)} \\
 & \textbf{$\delta$=-0.20} & \textbf{$\delta$=0.10} & \textbf{$\delta$=0.14} & \textbf{$\delta$=0.22} \\
\addlinespace
Gemini & 6.7 & \textbf{62$^{***}$} & \textbf{3,301$^{***}$} & \textbf{57k$^{*}$} \\
($n$=175) & (4.5--9.4) & \textbf{(30--139)} & \textbf{(1,274--9,688)} & \textbf{(15k--189k)} \\
 & $\delta$=0.01 & \textbf{$\delta$=0.19} & \textbf{$\delta$=0.19} & \textbf{$\delta$=0.11} \\
\addlinespace
Codex$^{****}$ & 7.6 & 46 & 6,014 & \textbf{505k$^{*}$} \\
($n$=4) & (4.6--11) & (32--62) & (4,997--6,514) & \textbf{(153k--939k)} \\
 & $\delta$=0.12 & $\delta$=-0.07 & $\delta$=0.48 & \textbf{$\delta$=0.66} \\
\bottomrule
\end{tabular}
\vskip 2pt
\begin{minipage}{\columnwidth}
\footnotesize
\textit{Note:} Mann--Whitney U test (two-sided) comparing each tool's adopters against non-adopters (complement), with Benjamini--Hochberg (BH) FDR correction (6 categories $\times$ 4 metrics = 24 comparisons). Effect sizes: Cliff's delta. $^{*}$ $p < 0.05$; $^{**}$ $p < 0.01$; $^{***}$ $p < 0.001$ (adjusted). $^{****}$ Interpret Codex results with caution due to sample size.
\end{minipage}
\end{table}

\section{Details of Configuration Mechanisms (RQ3)}
\label{sec:results_rq3}
We analyzed three mechanisms in detail: \textsc{Context Files} and \textsc{Skills} because they are supported by all tools, and \textsc{Subagents} because they follow a similar format as \textsc{Skills}.

\subsection{Configuration Mechanism: \textsc{Context Files}}
\textsc{Context Files} are Markdown files that provide machine-readable context about a project. \texttt{AGENTS.\allowbreak{}md}, initially introduced by OpenAI, serves as an open tool-agnostic convention with growing cross-tool support~\cite{Seyedmoein2026}.
We identified 4,768 \textsc{Context Files} across 2,586 of the 2,853 repositories in our sample (90.6\%). Most repositories include one or two such files (Figure~\ref{fig:files-per-repo}).

Of the 4,768 \textsc{Context Files}, \texttt{CLAUDE.\allowbreak{}md} is most common (1,640 files, 34.4\%), followed by \texttt{AGENTS.\allowbreak{}md} (1,508; 31.6\%) and \texttt{copilot-instructions.\allowbreak{}md} (1,393; 29.2\%). \texttt{GEMINI.\allowbreak{}md} (154 files, 3.2\%) and \texttt{.cursorrules} (73 files, 1.5\%) are rare (\texttt{.cursorrules} are now deprecated in favor of \texttt{AGENTS.\allowbreak{}md} and are distinct from Cursor's \textsc{Rules}, stored in \texttt{.\allowbreak{}cursor/\allowbreak{}rules/\allowbreak{}}). At the repository level, among the 2,586 repositories with \textsc{Context Files}, \texttt{CLAUDE.\allowbreak{}md} has the highest adoption rate at 45.9\% (1,187 repos), followed by \texttt{AGENTS.\allowbreak{}md} (39.5\%; 1,021 repos) and \texttt{copilot-instructions.\allowbreak{}md} (36.0\%; 930 repos).
\textsc{Context File} adoption was uniformly high across languages (88.5--96.2\%). \texttt{CLAUDE.\allowbreak{}md} was the most common \textsc{Context File} across languages, except for Java, C\#, and C++, where \texttt{copilot-instructions.\allowbreak{}md} dominated.

\begin{figure*}[t]
    \centering
    \includegraphics[width=0.8\textwidth]{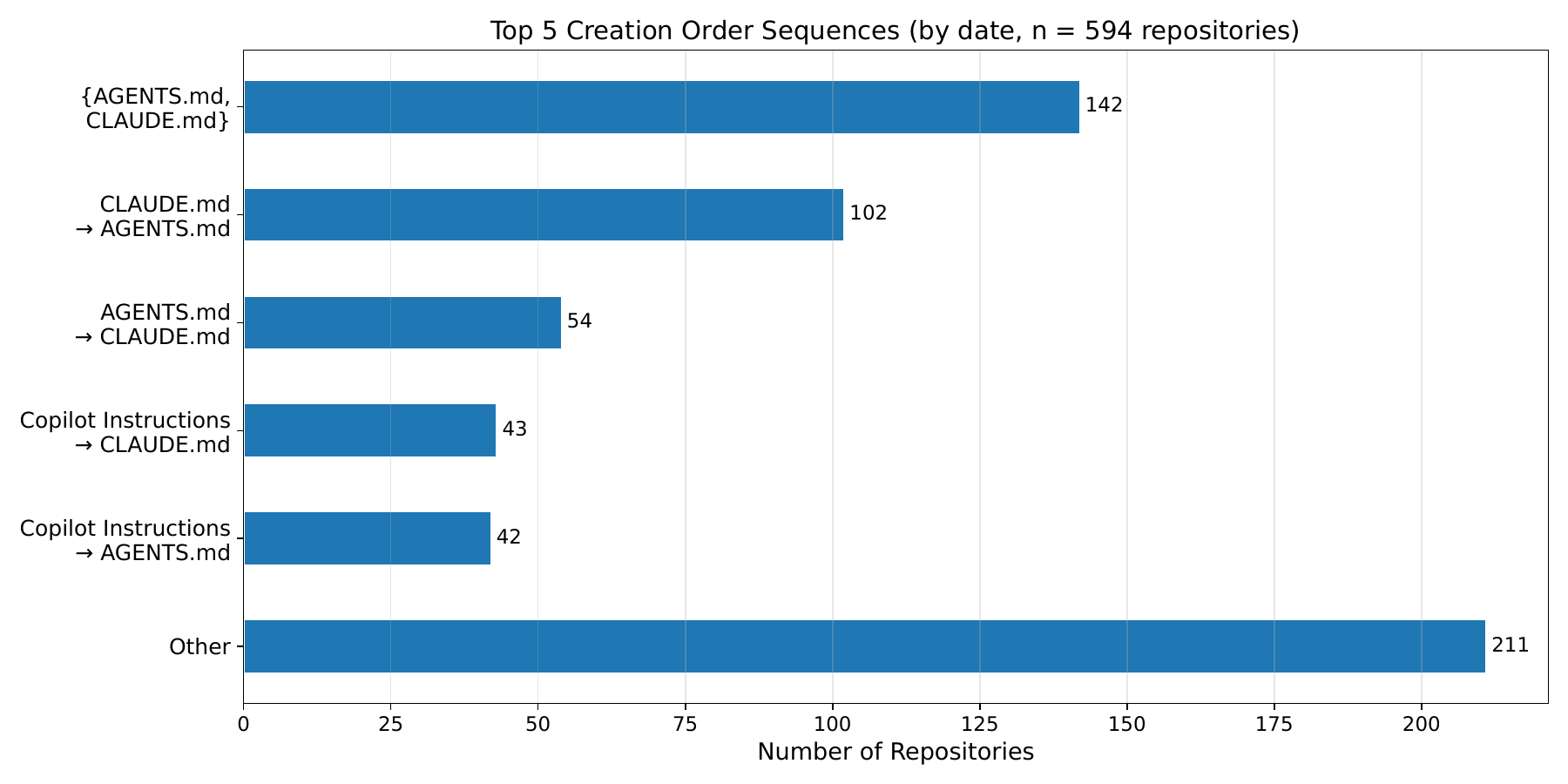}
    \caption{Creation order of \textsc{Context Files} per repository. Curly braces indicate that files were added on the same day.} 
    \Description{Diagram showing the creation order of Context Files per repository.}
    \label{fig:creation-order}
\end{figure*}

\begin{figure}[t]
    \centering
    \includegraphics[width=\columnwidth]{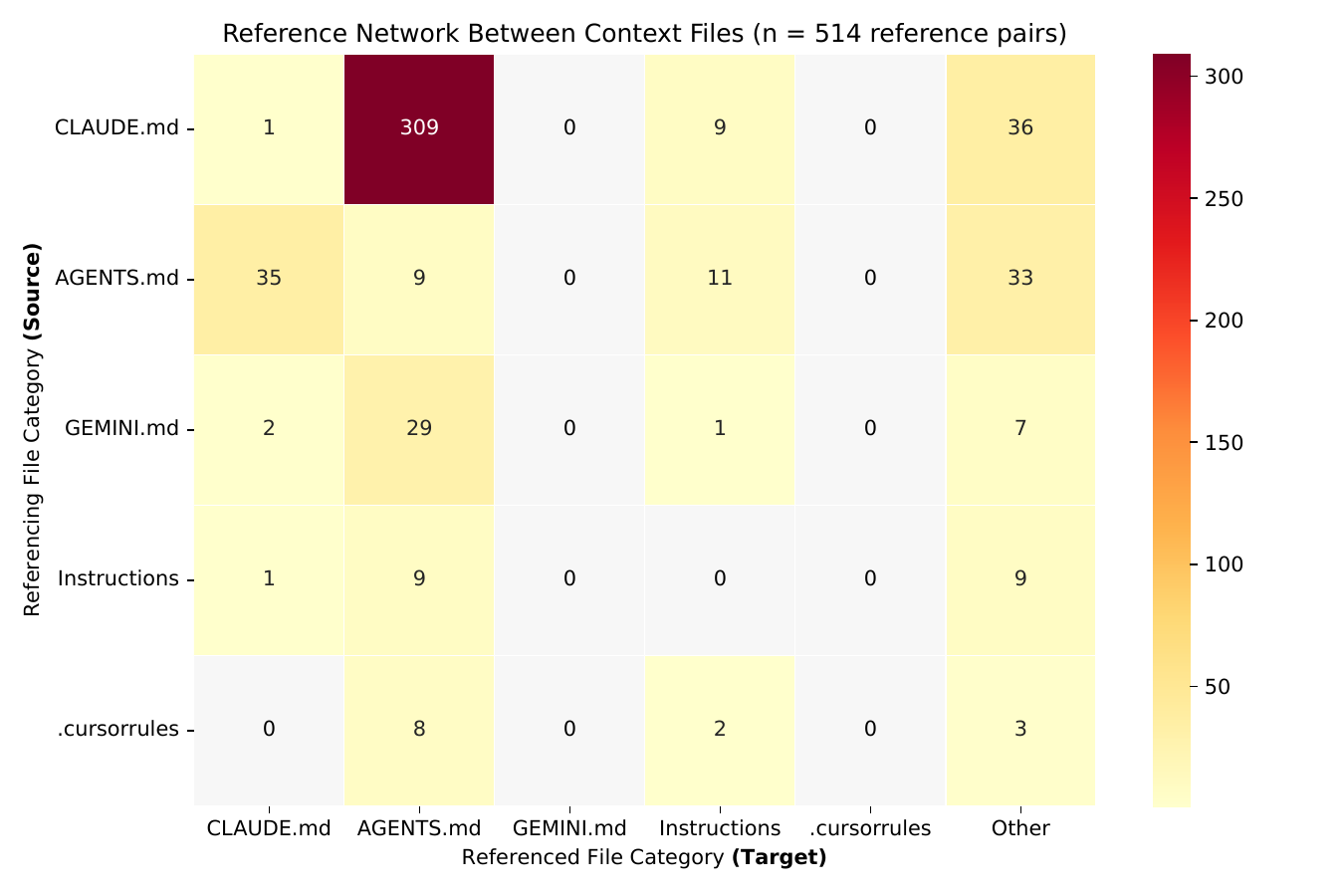}
    \caption{Reference-only \textsc{Context Files}.
    Examples of other files include \texttt{README.\allowbreak{}md} and \texttt{CONTRIBUTING.\allowbreak{}md}.}
    \Description{Diagram showing the reference patterns between Context Files.}
    \label{fig:references}
\end{figure}

Figure~\ref{fig:creation-order} shows the creation order of \textsc{Context Files} in repositories with multiple types (single-type repositories are omitted).
\texttt{CLAUDE.\allowbreak{}md} was typically created first, with \texttt{AGENTS.\allowbreak{}md} commonly added afterward---likely because Claude Code is the most popular agentic tool but does not yet support the emerging standard \texttt{AGENTS.\allowbreak{}md}~\cite{claude-code-agentsmd}.
Repositories that started with \texttt{copilot-instructions.\allowbreak{}md} often later added \texttt{CLAUDE.\allowbreak{}md} or \texttt{AGENTS.\allowbreak{}md}, despite Copilot supporting all major \textsc{Context File} types, reinforcing their de facto standard status.

Finally, \textsc{Context Files} commonly reference each other (Figure~\ref{fig:references}). We identified three types of references, where a source file points to a target file rather than providing its own content:

\begin{enumerate}
\item Direct pointer: \textsc{Context Files} include one line that references files either via their filename (e.g., \enq{\texttt{AGENTS.\allowbreak{}md}}, \enq{\texttt{@./\allowbreak{}AGENTS.md}}), imperative statements (e.g., \enq{Read \texttt{@AGENTS.md}}), or a Markdown link (e.g., \enq{\texttt{[anchor text](AGENTS.md)}}).

\item Short reference with context: \textsc{Context Files} provide 2--5 lines of what to do with the referenced file, e.g., 
\enq{Use instructions from \texttt{AGENTS.\allowbreak{}md} to guide your work.}, 
\enq{Always read \texttt{AGENTS.\allowbreak{}md} before answering}, or a  
Markdown header followed by one of the above.

\item Brief summary plus reference: \textsc{Context Files} include a  header, 1--2 sentences of context, and a reference to another file, e.g., \enq{\# AI Guidelines [...] **Read \texttt{@AGENTS.md} for comprehensive guidelines.**}

\end{enumerate}

We found 497 reference pairs in our sample. \texttt{CLAUDE.\allowbreak{}md} had the most outgoing references (344), predominantly pointing to \texttt{AGENTS.\allowbreak{}md} (301 times). Conversely, \texttt{AGENTS.\allowbreak{}md} received the most incoming references (353) and, when acting as a pointer, most frequently referenced \texttt{CLAUDE.\allowbreak{}md}; the same pairing pattern appears in the creation-order analysis.

\subsection{Configuration Mechanism: \textsc{Skills}}

\textsc{Skills} were introduced by Anthropic and now serve as an open standard for extending agentic AI tools with specialized contextual knowledge and workflows. \textsc{Skills} bundle prompts, tools, and documentation that an agent can invoke on demand. Each \textsc{Skill} is a directory containing a \texttt{SKILL.\allowbreak{}md} file with YAML frontmatter (\texttt{name}, \texttt{description}) and instructions in the body. The specification recommends that \texttt{SKILL.\allowbreak{}md} files contain fewer than 500 lines, with detailed material moved to separate files~\cite{agentskills}.
To provide detailed information and instructions, \textsc{Skills} can include additional resource directories loaded by agents when required:
\begin{enumerate}
\item \texttt{scripts/\allowbreak{}} contains executable code that agents can run.
Language options include Python, Bash, and JavaScript.
\item \texttt{references/\allowbreak{}} contains documentation that an agent can read when needed. This can include technical references, 
templates or structured data   
and other domain-specific files.
\item \texttt{assets/\allowbreak{}} contains static resources, such as templates (document templates, configuration templates), images (diagrams, examples) and data files (lookup tables, schemas).
\end{enumerate}

We found 601 \textsc{Skills} in 158 repositories. On average, these repositories use 3.8 \textsc{Skills} ($min = 1.0 $; $max = 28.0 $; $median = 2$). 
The distribution is right-skewed: most repositories define only one or two \textsc{Skills}, while a few define eight or more (see Figure~\ref{fig:files-per-repo}).
Only 29 of the 601 \textsc{Skills} (4.8\%) exceeded the 500-line recommendation.
Repositories defining \textsc{Skills} show the same language distribution as the overall agentic-tool sample, with TypeScript and Python being the most common (see Figure~\ref{fig:language-tool}).

We scanned all \textsc{Skills} directories for resource folders.
Of the 601 skills analyzed, the vast majority (514, 85.5\%) included no additional resources. Among those that did, the most common patterns were a \texttt{references} directory and a \texttt{scripts} directory (35 skills each, 5.8\%). Eleven \textsc{Skills} (1.8\%) combined both \texttt{scripts} and \texttt{references}, while an \texttt{assets} directory appeared in only four \textsc{Skills} (0.7\%). The remaining combinations, i.e., \texttt{references} with \texttt{assets}, and all three directories together, each occurred only once (0.2\%).
In summary, \textsc{Skills} mostly use static resources (documentation that the agent reads when needed) rather than dynamic resources, such as executable scripts that extend agent behavior. 

\subsection{Configuration Mechanism: \textsc{Subagents}}

\textsc{Subagents} are specialized AI agents to which agentic AI coding tools or other agents can delegate tasks. They share the same YAML frontmatter structure as \textsc{Skills} but operate within their own context window and return results to the parent agent, whereas \textsc{Skills} execute within the calling agent's context~\cite{claudeSubagents, cursorSubagents}.

We found 450 \textsc{Subagents} across 131 repositories. On average, these repositories use 3.44 \textsc{Subagents} ($min = 1.0$; $max = 17.0$; $median = 2$). As with \textsc{Skills}, the distribution is right-skewed: most repositories define only one or two \textsc{Subagents}, while a few define eight or more (Figure~\ref{fig:files-per-repo}). Claude Code's \textsc{Subagents} can have their own \emph{memory}, i.e., a persistent directory that survives across interactions and can build up knowledge over time, e.g., to store debugging insights~\cite{claudeSubagents}. However, we did not find repositories that store such memory files.

\begin{results}
\textbf{RQ3 (Summary):}
\begin{itemize}[leftmargin=*]
\item \textsc{Context~Files:} The most common artifacts were \texttt{CLAUDE.\allowbreak{}md}, \texttt{AGENTS.\allowbreak{}md} and \texttt{copilot-instructions.\allowbreak{}md}. 
File creation and reference patterns revealed that \texttt{CLAUDE.\allowbreak{}md} and \texttt{AGENTS.\allowbreak{}md} serve as de facto standards, with \texttt{AGENTS.\allowbreak{}md} becoming a tool-agnostic standard adopted across tools.

\item \textsc{Skills:} Despite their extensibility, most repositories that adopt \textsc{Skills} only define one or two and predominantly use static documentation rather than dynamic scripts. 

\item \textsc{Subagents:} Usage patterns match those of \textsc{Skills}, with most repositories defining only one or two. No repositories use the persistent memory feature available for Claude Code's \textsc{Subagents}.
\end{itemize}
\end{results}

\subsection{Threats to Validity}

We organize this section following established guidelines~\cite{DBLP:journals/ese/RunesonH09}.

\paragraph{Construct validity:}
Our heuristics (Table~\ref{tab:config-options-overview}) detect the \emph{presence} of configuration artifacts but not whether the corresponding tool is actively used, and a generic file such as \texttt{AGENTS.\allowbreak{}md} could in principle refer to any kind of AI agent, not only those used for software development.
Two sampling filters mitigate this threat: The language filter restricts repositories to the ten most common programming languages on GitHub (Python, TypeScript, JavaScript, Go, Java, C++, Rust, PHP, C\#, and~C) so that every repository in our sample is most likely a software project, and the ``engineered project'' classification further excludes, among others, tutorials and resource-aggregation repositories.
We additionally scanned the complete commit history of every repository that had at least one configuration file for AI-authored commits, using author/committer identity fields (verified bot accounts such as \texttt{copilot-swe-agent[bot]} and \texttt{claude[bot]}), git trailers (e.g., \texttt{Co-authored-by}), and commit message patterns. Among the 2,853 repositories with at least one configuration artifact, 2,058 (72.1\%) also contain AI-authored commits from these tools, a lower bound because commit detection requires explicit attribution that not every tool or workflow leaves behind. This provides direct evidence that detected artifacts are associated with active AI-coding tool usage. The detection scripts and full results are included in the supplementary material~\cite{supplementary-material}.

Additionally, for GitHub Copilot, Cursor, and Gemini, detected files apply to both conversational and agentic modes, so we cannot isolate agentic usage.
This limitation does not affect agentic-only mechanisms such as \textsc{Skills} and \textsc{Subagents}, nor the broader \texttt{AGENTS.\allowbreak{}md} trend.
Although multi-tool overlap could conflate tool-specific patterns, 85.4\% of tool-adopting repositories configure a single tool, limiting this concern.
To verify that detected artifacts contain meaningful content, we programmatically scanned all configuration files across the eight artifact types. We excluded empty files, files inside vendored directories (i.e., bundled copies of third-party dependencies, such as \texttt{node\_modules/\allowbreak{}} or \texttt{vendor/\allowbreak{}}), and files our earlier ``\texttt{.\allowbreak{}claude-plugin} parent'' rule had over-collected from nested submodules. This reduced the initial 2,926 detected repositories to 2,853 through two cleanup steps. The first step removed 45 repositories whose detected configurations were all discarded: 42 repositories whose only artifacts came from vendored dependencies 
and 3 removed for other reasons such as empty files.
The second step excluded a further 28 repositories whose only configuration files were in spec directories (e.g., \texttt{.\allowbreak{}claude/\allowbreak{}}, \texttt{.\allowbreak{}cursor/\allowbreak{}}) but did not match any of the eight named mechanisms or \texttt{AGENTS.\allowbreak{}md}.

Finally, one of the authors used Claude Code (Opus 4.6 with high effort) to check all 451 files with ten or fewer lines. Of these, 396 contained substantive configuration content (e.g., coding conventions, agent definitions, tool settings), 51 were reference-only files analyzed separately in Section~\ref{sec:results_rq3}, and four were borderline cases (e.g., a bare heading or filename). The supplementary material includes the inspection script, the checked files, and their classification.

\paragraph{Internal validity:}
We classified repositories as ``engineered'' software projects based on their \texttt{README} content. During iterative prompt development, we tested GPT-5-mini, GPT-5-nano, and GPT-5.2. We selected GPT-5.2 because it produced fewer ``unsure'' labels and spot-checks revealed that it produced fewer false positives and negatives.
We used a single labeling run; 2,204 ``unsure'' cases (often due to inaccessible linked resources) were excluded.
Future work should assess the reliability of this approach by adding alternative models and configurations.
Detection heuristics were designed by one author and cross-checked by two different authors.
Tool conventions evolve quickly; new mechanisms may not yet be captured.

\paragraph{External validity:}
Our study covers only open-source repositories on GitHub; practices in proprietary or enterprise settings may differ.
We selected repositories showing established engineering practices, but cannot claim representativeness for closed-source development, nor did we examine variation across application domains.
Finally, our findings are a point-in-time snapshot (February 2026) of a field whose conventions are still consolidating.

\section{Discussion}
\label{sec:discussion}

Our study offers a first cross-tool snapshot of configuration mechanisms and their artifacts across five agentic AI coding tools and 2,853 GitHub repositories. Below, we discuss the implications of our findings.
These mechanisms are a repository-level record of harness engineering. Developers widely adopt the static context mechanisms (e.g., \textsc{Context Files}) but rarely the executable and external ones (e.g., \textsc{Skills} with scripts, \textsc{Hooks}, and \textsc{MCP}). Harness engineering in open source today is therefore mostly context engineering.

\paragraph{Standardization around \texttt{AGENTS.\allowbreak{}md}:}
Our results show convergence on \texttt{AGENTS.\allowbreak{}md} as a tool-agnostic configuration file. Creation order analysis shows that \texttt{CLAUDE.\allowbreak{}md} typically appears first with \texttt{AGENTS.\allowbreak{}md} added later (Figure~\ref{fig:creation-order}). Reference patterns confirm this: \texttt{CLAUDE.\allowbreak{}md} most frequently points to \texttt{AGENTS.\allowbreak{}md} (301 cases), which receives 353 incoming references overall, far more than any other file type. Together, these patterns suggest bottom-up convergence on \texttt{AGENTS.\allowbreak{}md} across tools, driven by developer practice rather than vendor mandate. This is further reinforced by the 493 repositories that use \texttt{AGENTS.\allowbreak{}md} without any tool-specific configuration, indicating that developers adopt it as a tool-agnostic standard. At the same time, layering multiple context files in a single repository creates a risk of redundant or conflicting instructions. Native \texttt{AGENTS.\allowbreak{}md} support is already provided by Copilot, Cursor, and Codex; Claude Code, despite being the most popular tool, does not yet support it (Section~\ref{sec:results_rq3}), creating a clear gap between developer practice and tool capability.

\paragraph{Limited adoption of advanced mechanisms:}
Adoption of all three mechanisms remains limited. For each mechanism, most repositories include only one or two artifacts (Figure~\ref{fig:files-per-repo}). Moreover, 85.5\% of \textsc{Skills} do not include additional resources, and when resources are present, static documentation (\texttt{references/\allowbreak{}}) is as common as executable scripts (\texttt{scripts/\allowbreak{}}). \textsc{Skills} therefore function primarily as structured text rather than executable scripts. We also found no evidence of repositories using the persistent memory feature of Claude Code's \textsc{Subagents}.

This gap likely reflects both the novelty of these mechanisms and the effort required to configure them. Developers may prefer the simplest mechanism (i.e., \textsc{Context Files}), since defining executable \textsc{Skills} with scripts and structured resources requires more design and maintenance effort than authoring Markdown files. Future work should assess whether deeper configuration leads to measurable performance gains, extending early evidence on the impact of \textsc{Context Files}~\cite{lulla2026impact}.

\paragraph{Distinct tool ecosystems:}
Configuration practices differ across tools: Claude Code repositories use the broadest range of mechanisms, Cursor projects emphasize \textsc{Rules} and \textsc{Commands}, and Copilot repositories rarely extend beyond \textsc{Context Files}. These differences likely reflect the configuration options each tool exposes (see Table~\ref{tab:config-options-overview}). Repository characteristics also differ: Cursor repositories tend to be younger and larger, while Gemini repositories show higher activity levels (see Table~\ref{tab:repo-metadata-comparison}). It remains an open question whether these tool-specific configuration practices will converge as tool capabilities increasingly overlap, or whether distinct usage patterns will persist.

\paragraph{Practical implications and future directions:}
Table~\ref{tab:config-options-overview} helps developers understand available configuration mechanisms.
Based on our findings, we highlight four implications for practitioners.
First, \textsc{Context Files}, particularly \texttt{AGENTS.\allowbreak{}md}, are the simplest entry point for configuring agentic tools and are already widely adopted.
Second, developers who rely on multiple tools should maintain an \texttt{AGENTS.\allowbreak{}md} file as the shared core configuration, given its cross-tool support and the reference patterns we observed. When multiple \textsc{Context Files} coexist, teams may benefit from structuring them hierarchically (e.g., using tool-specific files as adapters that reference a shared core file).
Third, for recurring, well-defined workflows, \textsc{Skills} offer the option of bundling scripts and structured resources, but in practice most \textsc{Skills} do not include such resources.
Finally, adopting tool-specific mechanisms such as \textsc{Rules} or \textsc{Commands} ties workflows to a specific tool.

For researchers, our findings motivate longitudinal studies to track how configuration practices evolve as tools mature, and controlled studies to assess whether \textsc{Skills} with executable resources or dedicated \textsc{Subagents} provide measurable benefits over \textsc{Context Files} alone~\cite{lulla2026impact}.
Given that some repositories configure multiple tools (Section~\ref{subsec:characterization_of_repositories}), research should also investigate how conflicts between overlapping instructions across files can be detected and resolved.
Our data collection and analysis pipeline~\cite{supplementary-material} supports such ongoing analyses.

Finally, our findings represent a point-in-time snapshot.
The trends we identify---toward standardization around \texttt{AGENTS.\allowbreak{}md}, limited adoption of advanced mechanisms, and tool-specific configuration practices---are early empirical signals rather than settled findings.

\section{Conclusions}
\label{sec:conclusion}

Our study provides the first systematic, repository-level analysis of harness engineering for agentic AI coding tools, identifying eight configuration mechanisms across five tools and analyzing their adoption in 2,853 GitHub repositories.
Three findings stand out.
First, \textsc{Context Files} dominate and are often the only mechanism present. \texttt{AGENTS.\allowbreak{}md} is adopted across tools as a tool-agnostic standard, even when a developer's primary tool does not yet support it natively.
Second, few repositories adopt advanced mechanisms such as \textsc{Skills} and \textsc{Subagents}; most \textsc{Skills} do not bundle scripts or other executable resources, and we found no repositories using the persistent memory feature available for Claude Code's \textsc{Subagents}.
Third, distinct configuration practices are forming around different tools: Claude Code repositories use the broadest range of mechanisms, Cursor projects emphasize \textsc{Rules} and \textsc{Commands}, and Copilot repositories rarely extend beyond \textsc{Context Files}.

For practitioners, \texttt{AGENTS.\allowbreak{}md} is a natural starting point for configuring agentic tools, especially in multi-tool environments, and \textsc{Skills} offer further options for encoding recurring workflows through scripts and structured resources.
Tool providers should improve onboarding for \textsc{Skills} and \textsc{Subagents}, which few repositories use despite their support for executable scripts and isolated agent contexts.
For researchers, controlled studies should determine whether richer configuration improves outcomes over \textsc{Context Files} alone, and longitudinal research should track how these patterns evolve as tools mature and developer norms consolidate.

\balance
\bibliographystyle{ACM-Reference-Format}
\bibliography{literature}

@misc{Jiang2026,
  author        = {Shaokang Jiang and Daye Nam},
  title         = {Beyond the Prompt: An Empirical Study of {Cursor} Rules},
  year          = {2025},
  eprint        = {2512.18925},
  archivePrefix = {arXiv},
  primaryClass  = {cs.SE},
  doi           = {10.48550/arXiv.2512.18925},
  url           = {https://arxiv.org/abs/2512.18925},
  note          = {To appear at the 23rd IEEE/ACM International Conference on Mining Software Repositories (MSR 2026), Rio de Janeiro, Brazil}
}

@misc{He2026,
  author        = {Hao He and Courtney Miller and Shyam Agarwal and Christian K{\"a}stner and Bogdan Vasilescu},
  title         = {Speed at the Cost of Quality: How {Cursor} {AI} Increases Short-Term Velocity and Long-Term Complexity in Open-Source Projects},
  year          = {2025},
  eprint        = {2511.04427},
  archivePrefix = {arXiv},
  primaryClass  = {cs.SE},
  doi           = {10.48550/arXiv.2511.04427},
  url           = {https://arxiv.org/abs/2511.04427},
  note          = {To appear at the 23rd IEEE/ACM International Conference on Mining Software Repositories (MSR 2026), Rio de Janeiro, Brazil}
}

@article{10.1145/3715003,
author = {Terragni, Valerio and Vella, Annie and Roop, Partha and Blincoe, Kelly},
title = {The Future of {AI}-Driven Software Engineering},
year = {2025},
issue_date = {June 2025},
publisher = {Association for Computing Machinery},
address = {New York, NY, USA},
volume = {34},
number = {5},
issn = {1049-331X},
url = {https://doi.org/10.1145/3715003},
doi = {10.1145/3715003},
journal = {ACM Trans. Softw. Eng. Methodol.},
month = may,
articleno = {120},
numpages = {20}
}

@article{SERGEYUK2025107610,
title = {Using {AI}-based coding assistants in practice: State of affairs, perceptions, and ways forward},
journal = {Information and Software Technology},
volume = {178},
pages = {107610},
year = {2025},
issn = {0950-5849},
doi = {https://doi.org/10.1016/j.infsof.2024.107610},
url = {https://www.sciencedirect.com/science/article/pii/S0950584924002155},
author = {Agnia Sergeyuk and Yaroslav Golubev and Timofey Bryksin and Iftekhar Ahmed}
}

@misc{agentsmd_spec,
  title        = {{AGENTS.md}: A Simple, Open Format for Guiding Coding Agents},
  author       = {{agentsmd community}},
  year         = {2025},
  howpublished = {Website},
  url          = {https://agents.md/},
  note         = {Accessed 2026-01-18}
}

@misc{Seyedmoein2026,
  author        = {Seyedmoein Mohsenimofidi and Matthias Galster and Christoph Treude and Sebastian Baltes},
  title         = {Context Engineering for {AI} Agents in Open-Source Software},
  year          = {2025},
  eprint        = {2510.21413},
  archivePrefix = {arXiv},
  primaryClass  = {cs.SE},
  doi           = {10.48550/arXiv.2510.21413},
  url           = {https://arxiv.org/abs/2510.21413},
  note          = {To appear at the 23rd IEEE/ACM International Conference on Mining Software Repositories (MSR 2026), Rio de Janeiro, Brazil}
}

@misc{DBLP:journals/corr/abs-2507-13334,
  author        = {Lingrui Mei and
                   Jiayu Yao and
                   Yuyao Ge and
                   Yiwei Wang and
                   Baolong Bi and
                   Yujun Cai and
                   Jiazhi Liu and
                   Mingyu Li and
                   Zhong{-}Zhi Li and
                   Duzhen Zhang and
                   Chenlin Zhou and
                   Jiayi Mao and
                   Tianze Xia and
                   Jiafeng Guo and
                   Shenghua Liu},
  title         = {A Survey of Context Engineering for Large Language Models},
  year          = {2025},
  eprint        = {2507.13334},
  archivePrefix = {arXiv},
  primaryClass  = {cs.CL},
  doi           = {10.48550/arXiv.2507.13334},
  url           = {https://arxiv.org/abs/2507.13334}
}

@misc{DBLP:journals/corr/abs-2402-07927,
  author        = {Pranab Sahoo and
                   Ayush Kumar Singh and
                   Sriparna Saha and
                   Vinija Jain and
                   Samrat Mondal and
                   Aman Chadha},
  title         = {A Systematic Survey of Prompt Engineering in Large Language Models:
                   Techniques and Applications},
  year          = {2024},
  eprint        = {2402.07927},
  archivePrefix = {arXiv},
  primaryClass  = {cs.AI},
  doi           = {10.48550/arXiv.2402.07927},
  url           = {https://arxiv.org/abs/2402.07927}
}

@misc{philschmidSkillPrompting,
	author = {Philipp Schmid},
	title = {The New Skill in AI is Not Prompting, It's Context Engineering},
	howpublished = {\url{https://www.philschmid.de/context-engineering}},
    year= {2025},
	
}

@misc{dexterHorthyComplexCodebases,
	author = {Dexter Horthy},
	title = {Getting AI to Work in Complex Codebases},
	howpublished = {\url{https://github.com/humanlayer/advanced-context-engineering-for-coding-agents/blob/main/ace-fca.md}},
    year= {2025},
	
}

@misc{promptingguideElementsPrompt,
	author = {{DAIR.AI Prompt Engineering Guide}},
	title = {{E}lements of a {P}rompt | {P}rompt {E}ngineering {G}uide},
	howpublished = {\url{https://www.promptingguide.ai/introduction/elements}},
    year= {2025},
	
}

@misc{openaiIntroducingCodex,
	author = {{OpenAI}},
	title = {{I}ntroducing {C}odex},
	howpublished = {\url{https://openai.com/index/introducing-codex/}},
	year= {2025},
}

@misc{seartGitHubSearch,
	author = {{SEART}},
	title = {GitHub Search},
	howpublished = {\url{https://seart-ghs.si.usi.ch/}},
	year = {2025},
}

@misc{agentskills,
	author = {{agentskills.io}},
	title = {Agent Skills},
	howpublished = {\url{https://agentskills.io/}},
	year = {2026},
}

@misc{cursorSubagents,
	author = {{Cursor}},
	title = {{S}ubagents},
	howpublished = {\url{https://cursor.com/docs/context/subagents}},
	year = {2026},
}

@misc{claudeSubagents,
	author = {Anthropic},
	title = {{C}reate custom subagents},
	howpublished = {\url{https://code.claude.com/docs/en/sub-agents}},
	year = {2026},
}

@misc{anthropicClaudeCodeRelease,
	author = {{Anthropic}},
	title = {Claude 3.7 Sonnet and Claude Code},
	howpublished = {\url{https://www.anthropic.com/news/claude-3-7-sonnet}},
	year = {2025},
}

@article{DBLP:journals/ese/MunaiahKCN17,
  author       = {Nuthan Munaiah and
                  Steven Kroh and
                  Craig Cabrey and
                  Meiyappan Nagappan},
  title        = {Curating {GitHub} for engineered software projects},
  journal      = {Empir. Softw. Eng.},
  volume       = {22},
  number       = {6},
  pages        = {3219--3253},
  year         = {2017},
  url          = {https://doi.org/10.1007/s10664-017-9512-6},
  doi          = {10.1007/S10664-017-9512-6},
  bibsource    = {dblp computer science bibliography, https://dblp.org}
}

@inproceedings{DBLP:conf/msr/DabicAB21,
  author       = {Ozren Dabic and
                  Emad Aghajani and
                  Gabriele Bavota},
  title        = {Sampling Projects in {GitHub} for {MSR} Studies},
  booktitle    = {18th {IEEE/ACM} International Conference on Mining Software Repositories,
                  {MSR} 2021, Madrid, Spain, May 17-19, 2021},
  pages        = {560--564},
  publisher    = {{IEEE}},
  address      = {Madrid, Spain},
  year         = {2021},
  url          = {https://doi.org/10.1109/MSR52588.2021.00074},
  doi          = {10.1109/MSR52588.2021.00074},
  bibsource    = {dblp computer science bibliography, https://dblp.org}
}

@misc{santos2025decoding,
  author        = {Helio Victor F. Santos and
                   Vitor Costa and
                   Jo{\~{a}}o Eduardo Montandon and
                   Marco T{\'{u}}lio Valente},
  title         = {Decoding the Configuration of {AI} Coding Agents: Insights from {Claude Code} Projects},
  year          = {2025},
  eprint        = {2511.09268},
  archivePrefix = {arXiv},
  primaryClass  = {cs.SE},
  doi           = {10.48550/arXiv.2511.09268},
  url           = {https://arxiv.org/abs/2511.09268}
}

@misc{DBLP:journals/corr/abs-2511-12884,
  author        = {Worawalan Chatlatanagulchai and
                   Hao Li and
                   Yutaro Kashiwa and
                   Brittany Reid and
                   Kundjanasith Thonglek and
                   Pattara Leelaprute and
                   Arnon Rungsawang and
                   Bundit Manaskasemsak and
                   Bram Adams and
                   Ahmed E. Hassan and
                   Hajimu Iida},
  title         = {Agent {READMEs}: An Empirical Study of Context Files for Agentic Coding},
  year          = {2025},
  eprint        = {2511.12884},
  archivePrefix = {arXiv},
  primaryClass  = {cs.SE},
  doi           = {10.48550/arXiv.2511.12884},
  url           = {https://arxiv.org/abs/2511.12884}
}

@inproceedings{villamizar2025prompts,
  author       = {Hugo Villamizar and
                  Jannik Fischbach and
                  Alexander Korn and
                  Andreas Vogelsang and
                  Daniel M{\'{e}}ndez},
  title        = {Prompts as Software Engineering Artifacts: {A} Research Agenda and
                  Preliminary Findings},
  booktitle    = {Product-Focused Software Process Improvement - 26th International
                  Conference, {PROFES} 2025, Salerno, Italy, December 1-3, 2025, Proceedings},
  series       = {Lecture Notes in Computer Science},
  pages        = {470--478},
  publisher    = {Springer},
  address      = {Cham},
  year         = {2025},
  doi          = {10.1007/978-3-032-12089-2\_32}
}

@misc{mcmillan2026structured,
  author        = {Damon McMillan},
  title         = {Structured Context Engineering for File-Native Agentic Systems: Evaluating
                   Schema Accuracy, Format Effectiveness, and Multi-File Navigation at Scale},
  year          = {2026},
  eprint        = {2602.05447},
  archivePrefix = {arXiv},
  primaryClass  = {cs.SE},
  doi           = {10.48550/arXiv.2602.05447},
  url           = {https://arxiv.org/abs/2602.05447}
}

@misc{zhang2025agentic,
  author        = {Qizheng Zhang and
                   Changran Hu and
                   Shubhangi Upasani and
                   Boyuan Ma and
                   Fenglu Hong and
                   Vamsidhar Kamanuru and
                   Jay Rainton and
                   Chen Wu and
                   Mengmeng Ji and
                   Hanchen Li and
                   Urmish Thakker and
                   James Zou and
                   Kunle Olukotun},
  title         = {Agentic Context Engineering: Evolving Contexts for Self-Improving
                   Language Models},
  year          = {2025},
  eprint        = {2510.04618},
  archivePrefix = {arXiv},
  primaryClass  = {cs.CL},
  doi           = {10.48550/arXiv.2510.04618},
  url           = {https://arxiv.org/abs/2510.04618}
}

@misc{lulla2026impact,
  author        = {Jai Lal Lulla and
                   Seyedmoein Mohsenimofidi and
                   Matthias Galster and
                   Jie M. Zhang and
                   Sebastian Baltes and
                   Christoph Treude},
  title         = {On the Impact of {AGENTS.md} Files on the Efficiency of {AI} Coding Agents},
  year          = {2026},
  eprint        = {2601.20404},
  archivePrefix = {arXiv},
  primaryClass  = {cs.SE},
  doi           = {10.48550/arXiv.2601.20404},
  url           = {https://arxiv.org/abs/2601.20404},
  note          = {To appear at the 1st Journal Ahead Workshop ({JAWs}@{ICSE} 2026)}
}

@inproceedings{yang2024swe,
  author       = {John Yang and
                  Carlos E. Jimenez and
                  Alexander Wettig and
                  Kilian Lieret and
                  Shunyu Yao and
                  Karthik Narasimhan and
                  Ofir Press},
  title        = {{SWE-agent}: Agent-Computer Interfaces Enable Automated Software Engineering},
  booktitle    = {Advances in Neural Information Processing Systems 38: Annual Conference
                  on Neural Information Processing Systems 2024, {NeurIPS} 2024, Vancouver,
                  BC, Canada, December 10-15, 2024},
  pages        = {50528--50652},
  publisher    = {Curran Associates, Inc.},
  address      = {Red Hook, NY, USA},
  year         = {2024},
  doi          = {10.52202/079017-1601},
  url          = {http://papers.nips.cc/paper\_files/paper/2024/hash/5a7c947568c1b1328ccc5230172e1e7c-Abstract-Conference.html}
}

@misc{zhang2025monadic,
  author        = {Yifan Zhang and
                   Yang Yuan and
                   Mengdi Wang and
                   Andrew Chi-Chih Yao},
  title         = {Monadic Context Engineering},
  year          = {2025},
  eprint        = {2512.22431},
  archivePrefix = {arXiv},
  primaryClass  = {cs.AI},
  doi           = {10.48550/arXiv.2512.22431},
  url           = {https://arxiv.org/abs/2512.22431}
}

@misc{ye2026meta,
  author        = {Haoran Ye and
                   Xuning He and
                   Vincent Arak and
                   Haonan Dong and
                   Guojie Song},
  title         = {Meta Context Engineering via Agentic Skill Evolution},
  year          = {2026},
  eprint        = {2601.21557},
  archivePrefix = {arXiv},
  primaryClass  = {cs.AI},
  doi           = {10.48550/arXiv.2601.21557},
  url           = {https://arxiv.org/abs/2601.21557}
}

@dataset{supplementary-material,
  author       = {Galster, Matthias and Mohsenimofidi, Seyedmoein and Lulla, Jai Lal and Abubakar, Muhammad Auwal and Treude, Christoph and Baltes, Sebastian},
  title        = {Configuring Agentic AI Coding Tools: An Exploratory Study (Supplementary Material)},
  month        = mar,
  year         = 2026,
  publisher    = {Zenodo},
  doi          = {10.5281/zenodo.18625980},
  url          = {https://doi.org/10.5281/zenodo.18625980},
}

@inproceedings{galster2026configuring,
  author       = {Galster, Matthias and Mohsenimofidi, Seyedmoein and Lulla, Jai Lal and Abubakar, Muhammad Auwal and Treude, Christoph and Baltes, Sebastian},
  title        = {Configuring Agentic {AI} Coding Tools: An Exploratory Study},
  booktitle    = {Proceedings of the 3rd ACM International Conference on AI-Powered Software (AIware '26)},
  year         = {2026},
  publisher    = {ACM},
  address      = {Montreal, QC, Canada},
  isbn         = {979-8-4007-2601-9},
  doi          = {10.1145/3805760.3814887},
  url          = {https://doi.org/10.1145/3805760.3814887},
}

@misc{stackoverflow-survey-2025,
    title = {{Stack Overflow Developer Survey 2025: AI Agent out-of-the-box tools}},
    author = {{Stack Exchange Inc.}},
    year = {2026},
    howpublished = {\url{https://survey.stackoverflow.co/2025/ai/\#3-ai-agent-out-of-the-box-tools}}
}

@misc{claude-code-agentsmd,
    title = {{Feature Request: Support AGENTS.md}},
    author = {{anthropics/claude-code on GitHub}},
    year = {2026},
    howpublished = {\url{https://github.com/anthropics/claude-code/issues/6235}}
}

@article{DBLP:journals/ese/RunesonH09,
  author       = {Per Runeson and
                  Martin H{\"{o}}st},
  title        = {Guidelines for conducting and reporting case study research in software
                  engineering},
  journal      = {Empir. Softw. Eng.},
  volume       = {14},
  number       = {2},
  pages        = {131--164},
  year         = {2009},
  url          = {https://doi.org/10.1007/s10664-008-9102-8},
  doi          = {10.1007/S10664-008-9102-8},
  bibsource    = {dblp computer science bibliography, https://dblp.org}
}

@misc{DBLP:journals/corr/abs-2601-18341,
  author        = {Romain Robbes and
                   Th{\'{e}}o Matricon and
                   Thomas Degueule and
                   Andr{\'{e}} C. Hora and
                   Stefano Zacchiroli},
  title         = {Agentic Much? Adoption of Coding Agents on GitHub},
  year          = {2026},
  eprint        = {2601.18341},
  archivePrefix = {arXiv},
  primaryClass  = {cs.SE},
  doi           = {10.48550/arXiv.2601.18341},
  url           = {https://arxiv.org/abs/2601.18341}
}

@inproceedings{DBLP:conf/msr/KalliamvakouGBSGD14,
  author       = {Eirini Kalliamvakou and
                  Georgios Gousios and
                  Kelly Blincoe and
                  Leif Singer and
                  Daniel M. Germ{\'{a}}n and
                  Daniela E. Damian},
  title        = {The promises and perils of mining GitHub},
  booktitle    = {11th Working Conference on Mining Software Repositories, {MSR} 2014},
  pages        = {92--101},
  publisher    = {{ACM}},
  address      = {Hyderabad, India},
  year         = {2014},
  url          = {https://doi.org/10.1145/2597073.2597074},
  doi          = {10.1145/2597073.2597074}
}

@misc{DBLP:journals/corr/abs-2604-08224,
  author        = {Chenyu Zhou and
                   Huacan Chai and
                   Wenteng Chen and
                   Zihan Guo and
                   Rong Shan and
                   Yuanyi Song and
                   Tianyi Xu and
                   Yingxuan Yang and
                   Aofan Yu and
                   Weiming Zhang and
                   Congming Zheng and
                   Jiachen Zhu and
                   Zeyu Zheng and
                   Zhuosheng Zhang and
                   Xingyu Lou and
                   Changwang Zhang and
                   Zhihui Fu and
                   Jun Wang and
                   Weiwen Liu and
                   Jianghao Lin and
                   Weinan Zhang},
  title         = {Externalization in {LLM} Agents: {A} Unified Review of Memory, Skills, Protocols and Harness Engineering},
  year          = {2026},
  eprint        = {2604.08224},
  archivePrefix = {arXiv},
  primaryClass  = {cs.SE},
  doi           = {10.48550/arXiv.2604.08224},
  url           = {https://arxiv.org/abs/2604.08224}
}

\end{document}